\documentclass[12pt]{article}
\usepackage{sectsty}
\usepackage{mathrsfs}
\allsectionsfont{\sffamily}
\usepackage[nosort]{cite}

\usepackage[pdftex]{graphicx}
\usepackage[font={scriptsize,singlespacing}]{caption}
\usepackage{epstopdf}

\usepackage{amsmath}
\usepackage{amssymb}
\usepackage{subfigure}
\usepackage{hyperref}
\usepackage{url}
\usepackage{xcolor}
\usepackage{color}
\definecolor{amaranth}{rgb}{0.9, 0.17, 0.31}
\definecolor{purple(munsell)}{rgb}{0.62, 0.0, 0.77}
\definecolor{americanrose}{rgb}{1.0, 0.01, 0.24}
\definecolor{palatinateblue}{rgb}{0.15, 0.23, 0.89}
\definecolor{royalblue(web)}{rgb}{0.25, 0.41, 0.88}
\definecolor{hanpurple}{rgb}{0.32, 0.09, 0.98}
\definecolor{beaublue}{rgb}{0.74, 0.83, 0.9}
\definecolor{carminered}{rgb}{1.0, 0.0, 0.22}
\definecolor{brightpink}{rgb}{1.0, 0.0, 0.5}
\hypersetup{ linktoc=all,
    colorlinks, linkcolor={palatinateblue},
    citecolor={brightpink}, urlcolor={purple(munsell)}
}

\def\sideremark#1{\ifvmode\leavevmode\fi\vadjust{\vbox to0pt{\vss
 \hbox to 0pt{\hskip\hsize\hskip1em
 \vbox{\hsize2cm\tiny\raggedright\pretolerance10000
 \noindent #1\hfill}\hss}\vbox to8pt{\vfil}\vss}}}%
\newcommand{\bo}{\raise-1mm\hbox{\Large$\Box$}}

\newcommand{\be}{\begin{equation}}
\newcommand{\ee}{\end{equation}}
\newcommand{\bea}{\begin{eqnarray}}
\newcommand{\eea}{\end{eqnarray}}

\begin{document}
\thispagestyle{empty}
\begin{center}

\null \vskip-1truecm \vskip2truecm

{\Large{\bf \textsf{Black Hole Information Loss:\\ \vskip0.2truecm Some Food for Thoughts}}}

\vskip1truecm

{{Yen Chin Ong}}\\
\vskip0.1truecm
{Nordita, KTH Royal Institute of Technology and Stockholm University, \\ 
Roslagstullsbacken 23,
SE-106 91 Stockholm, Sweden}\\
{\tt Email: yenchin.ong@nordita.org}\\

\vskip0.6truecm

{{Dong-han Yeom}}\\
{{Leung Center for Cosmology and Particle Astrophysics,\\
National Taiwan University, Taipei 10617, Taiwan}\\
{\tt Email: innocent.yeom@gmail.com}}\\

\end{center}
\vskip1truecm \centerline{\textsf{ABSTRACT}} \baselineskip=15pt

\medskip
We provide some thoughts on the black hole information loss paradox, highlighting some important aspects of the problem that need to be addressed in order to resolve the paradox. 
\begin{quote}
{\tt This is a modified and elaborated version of the summary talk for the \emph{Black Hole Evaporation and Information Loss Paradox} parallel session, organized by the two of us, during the Second LeCosPA\newline Symposium \emph{Everything about Gravity}, which was held in National \newline Taiwan University, Taipei, in December 2015. }
\end{quote}

\section{The Usual Assumptions}

The information loss paradox \cite{Hawking:1976ra} remains an important, unresolved, problem in theoretical physics. For 40 years now, the literature is filled with countless proposals, ranging from the mundane to the preposterous, that claim to have -- at least partially -- resolve the paradox. However, not only was there no consensus whatsoever regarding any of these purported solutions, the mysteries deepened even further when the firewall\footnote{See also \cite{sam} and p.26 of \cite{giddings}.} controversy \cite{Almheiri:2012rt} started a few years ago.

In this note, we shall discuss some aspects of the information loss paradox, that we feel \emph{could} be important to clarify before we can fully resolve the paradox.

\newpage

As a reference we shall list the usual assumptions of black hole complementarity \cite{Susskind:1993if}:
\begin{itemize}
\item[$(1)$] \textit{Unitarity} holds throughout the entire process of black hole evaporation;
\smallskip
\item[$(2)$] \textit{No drama} for an in-falling observer in accordance to general relativity;
\smallskip
\item[$(3)$] \textit{Local quantum field theory} is correct, at least outside the stretched horizon of a sufficiently massive black hole;
\smallskip
\item[$(4)$] \textit{The statistical entropy of a black hole is proportional to the black hole area}: $\log N = \mathcal{A}/4$, where $N$ is the number of states of a black hole and $\mathcal{A}$ is the area of the black hole horizon;
\end{itemize}
\smallskip
To this we add an often implicit assumption that:
\begin{itemize}
\item[$(5)$] \textit{There is an observer} who can extract information from the Hawking radiation\footnote{This observer does not have to be an asymptotic observer. In view of the firewall proposal, We have in mind an observer whose role is to collect and attempt to decode the highly scrambled information encoded in the Hawking radiation\cite{HH}.}.
\end{itemize}
If we accept the various arguments -- such as the firewall claim -- that black hole complementarity is inconsistent \cite{Almheiri:2012rt,Yeom:2009zp}, then we need to drop at least one of these assumptions. Therefore, we have the following options (We denote the negation of the aforementioned assumptions with a ``$\neg$'' sign):
\begin{itemize}
\item[$\neg (1)$] ``Non-unitarity'' is not a serious problem: it need not violate energy-momentum conservation \cite{Unruh:1995gn}. It could also simply mean that there are ``extra places'' the information could go to, such as the singularity or a baby universe. 
\smallskip
\item[$\neg (2)$] Either general relativity breaks down not only at the singularity but at a larger scale and gives rise to new effects at the horizon, or that when the backreaction of Hawking radiation is properly taken into account (such as in the approach of \cite{Ho:2015vga}; see also the references therein) the horizon never formed\footnote{However, Unruh emphasized that black holes should form in his plenary talk ``Firewalls: What and Why the Issue''.}, or that no-drama is violated by the presence of a firewall. The firewall view has been criticized by many authors in the literature, for a recent example, see, e.g., \cite{Chen:2015gux}.
\smallskip
\item[$\neg (3)$] Local quantum field theory is violated, e.g., there may be non-local quantum effects \cite{Page:2013mqa, 1211.7070v3}.
\smallskip
\item[$\neg (4)$] $\log N \neq \mathcal{A}/4$. Regarding this possibility, there are various ways to realize the idea \cite{Chen:2014jwq}. For example,
\begin{itemize}
\item[(i)] Information can be compressed by the quantum gravitational regime, i.e. the entropy bound may be violated at late time\footnote{A point emphasized by Ashtekar in his talk ``Quantum Gravity: An Overview'', in another parallel session ``Emergent Gravity and Quantum Gravity''.}. See also \cite{Ashtekar:2005cj}. Hence, the resolution of the singularity may be sufficient, or perhaps \emph{essential} to explain the information loss problem.
\smallskip
\item[(ii)] A small (perhaps Planck scale) remnant can store all the information.
\smallskip
\item[(iii)] The interior of a black hole may store sufficiently large amount of information, e.g., in a large volume, similar to the ``bag-of-gold'' geometry, or a bubble universe.
\end{itemize}
\smallskip
\item[$\neg (5)$] There is no observer who can read information from the Hawking radiation; information does not disappear but semi-classical observers are not able to read the information \cite{Sasaki:2014spa,Chen:2015lbp,Chen:2015ibc}.
\end{itemize}

We shall now discuss some of these possibilities in more detail.

\section{Questions Regarding the Black Hole Geometry}
It is best to have in mind some crucial questions while one attempts to resolve the information loss paradox. The first important aspect is about the geometry of a black hole spacetime -- more specifically, the roles played by the horizon, the singularity, and the interior spacetime of a black hole. Let us elaborate:

\begin{itemize}
\item[(1)] \emph{Horizon:} In general relativity, a black hole is defined by its event horizon, beyond which nothing, not even light, can escape from. Other notions of horizons, such as apparent/trapping horizons, have also been discussed in the recent literature. An important and crucial question is whether a black hole, defined in the strict sense by the event horizon, really exist? Apparent horizons are not without problems since they are hypersurface dependent. (A better alternative may be the ``prison'', proposed by Don Page\cite{priv}, which for a fixed classical spacetime, is independent of the choice of slicing; it is the set of all points through which there is a trapped or
marginally trapped surface. The ``prison'', however, is much harder to calculate.) There exist attempts to prevent the information loss issue from ever arising, by arguing that event horizons do not form (and in some of these approaches, not even apparent horizons exist, see, e.g., \cite{Ho:2015vga}). 
\smallskip
\item[(2)] \emph{Singularity:} One often assumes that singularities in general relativity can be cured by quantum gravity. There are two issues here worth emphasizing: firstly, as emphasized by Unruh is his plenary talk, we do not have any evidence, other than wishful thinking, that the singularity can indeed be cured. If quantum gravity fails to cure the singularity\cite{Chen:2014jwq}, then perhaps information just falls through the edge of spacetime into the singularity. This possibility was emphasized in the plenary talk by Unruh. Even if quantum gravity successfully cures the singularity, it might be that the information simply falls into whatever it is that replaces the singularity. This is related to the second point we wish to raise here: can the information loss paradox be understood without the full knowledge of quantum gravity (including the full understanding of singularities)?
\smallskip
\item[(3)] \emph{Interior Spacetime:} The information loss paradox is often discussed from the point of view of the exterior observers. Perhaps some insights can be gained by investigating the issue from the point of view of an infalling observer instead. Does the dynamical interior of the black hole play any role at all in somehow ``storing the information''? Such a question is interesting considering the fact that in some sense the black hole interior spacetime has a very large spatial volume \cite{Ong:2015tua,1411.2854,1502.01907,1503.08245}, although a recent investigation suggests that the associated entropy of the large volume is only proportional to the surface area \cite{1510.02182}. A related issue, as raised by Unruh in his plenary talk, is: how do we understand the ``bag-of-gold'' type of geometry, especially in the context of AdS/CFT correspondence? That is, can the boundary probe the interior of the large volume of such a nontrivial geometry? This question\cite{0810.4886} was previously also discussed in the literature in the context of the so-called ``monster'' spacetime\cite{0706.3239,0908.1265,1304.3803}, but there is still some room for further study. 

For a more detailed discussion on points No.(2) and No.(3) above, see \cite{yco}. 

\end{itemize}

\section{What Counts as Resolving the Paradox?}

More fundamentally, there are some rather philosophical questions that are worth emphasizing. The most important issue is: \emph{what do we mean by resolving the information loss paradox?} The ``preferred'' approach in the literature seems to be that the \emph{external} observers should be able to, at least in principle, recover all the information after the black hole has evaporated away. The alternative solutions, such as having the information somehow locked inside a remnant \cite{Chen:2014jwq} or goes into a baby universe (or simply falls into the singularity if one takes general relativity at -- perhaps naive -- face value), remain somewhat unpopular. 

Note that one is usually not as bothered by, say, information from the very early cosmos that in principle could leak into the various bubble universes in the eternal inflationary scenario. 
Of course, this does not necessarily imply that there is a double standard: perhaps people who believe that information should not escape into a bubble universe would also have the same belief in the context of cosmology.  Specifically, information should not leak into bubble universes if they disconnected from the parent spacetime.  Such disconnected bubble universes would presumably have a unique initial quantum state. However,  in eternal inflationary scenarios, the bubbles expand at the speed of light and do not disconnect from the parent. This means that a growing part of the parent spacetime will be eaten up
by the bubbles (though what is left outside the bubbles may grow even faster, so that the growing bubbles do not eat up all or even most of the parent spacetime). One then has to include the information inside the bubbles as well as outside, because then any complete Cauchy-like surface to the future of the bubble formation would have to go inside the bubbles, whereas
if the bubbles disconnected, there would be Cauchy-like surfaces to
the future of the location of the bubble formation and disappearance
from the parent spacetime.  (If the bubble formation and disappearance
were modeled by a spacetime with a naked singularity there, a
Cauchy-like surface to the future would not really be a Cauchy
surface, but one might suppose that quantum effects heal the naked
singularity so that the quantum state on the Cauchy-like surface to
the future is a unitary transformation of the Cauchy-like surface to
the past of the bubble formation.)\footnote{We thank Don Page for his comments that resulted in this paragraph.}

Another equally important question which is related to the above discussion is: \emph{what do we really mean by unitarity?} As pointed out in quite a few talks in this parallel session \cite{Sasaki:2014spa, Chen:2015lbp, Chen:2015ibc}, perhaps we are still thinking ``too classically.'' (See also, \cite{nvw,1302.0451}.) That is, information loss only occurs if one looks at semi-classical picture in which matter is quantized but there is still one definite -- albeit evolving -- spacetime geometry. It is possible that all that is required for unitarity is that the information is conserved at the level of wavefunctions, and that we should instead consider superpositions of different geometries (and topologies). In such a picture, information is loss as far as semi-classical observers (who only see one definite geometry) are concerned\cite{1302.0451}, but unitarity remains intact at the fundamental level.

Lastly, although one wishes to find \emph{the} resolution of the information loss paradox, one should keep in mind the possibility that perhaps there is \emph{no unique resolution}. Afterall, there exist many types of black holes with various topologies and hairs. One often finds that a proposed resolution works for one type of black hole but not for another, more complicated, type. For example, even if the remnant proposal may overcome the usual challenges such as the infinite production problem (see \cite{Chen:2014jwq} for a review), one may need to worry that remnant itself is not stable, at least on certain backgrounds \cite{Wen:2015kwa}. Consider, as another example, the idea that the random walk motion of black holes, which is due to the backreaction from Hawking radiation emission, could resolve the information loss paradox if we subscribe to unitarity in the sense of the entire wavefunction (since among the superpositions, there exist some ``worlds'' in which a piece of information does not fall into the black hole due to the fact that the black hole has already moved out of the way) \cite{nvw,1302.0451}. While this may work in asymptotically flat spacetime, it fails in asymptotically anti-de Sitter spacetimes since the random walk motion is strongly suppresed in that case \cite{ampss}.

Perhaps Nature does respect unitarity and information is always conserved, but there are many ways to ensure this, instead of a simple elegant mean. 
Furthermore, even a specific type of black hole can be formed from many different initial conditions. As Stephen Fulling pointed out in his talk during the Hawking Radiation Conference held in Stockholm in August 2015, perhaps different initial conditions lead to black holes with different end states. For example, some might have a ``bag-of-gold'' while some have relatively ``trivial'' interiors. It is possible that some black holes might completely evaporate but some remain as a remnant.

The information loss paradox is indeed a long-standing problem in black hole physics. It is quite unfortunate that we have not made much progress in this subject after 40 years.
The firewall controversy demonstrates that our understanding of black hole physics is even less than previously thought. In view of this, hidden assumptions are starting to be examined, and further understanding would certainly be made. In addition, there are various other ways that may provide some insights into Hawking evaporation in general, such as the moving mirror approaches \cite{Good:2015nja, Good:2013lca, Good:2015jwa, Anderson:2015iga}, which can hopefully be tested experimentally in the near future \cite{Chen:2015bcg}. (A further understanding of the Unruh radiation \cite{Oshita:2015qka} may also help.) Hopefully by the 200th anniversary of general relativity, we would have the answer ready.

\addtocounter{section}{4}
\section*{\large{\textsf{Acknowledgement}}}

We thank all the speakers of the parallel sessions: Pei-Ming Ho, Wen-Yu Wen, Michael R. R. Good, Guillem Dom\'enech, Naritaka Oshita, and Yao-Chieh Hu. In addition we wish to thank Abhay Ashtekar, Pisin Chen, Don N. Page, and William G. Unruh, for useful discussions and comments.




\end{document}